\documentclass[journal]{IEEEtran}

\usepackage{cite}
\usepackage{enumitem}
\usepackage{footnote}
\usepackage{threeparttable}
\usepackage[table]{xcolor}
\usepackage{mathtools}

\ifCLASSINFOpdf
 \usepackage[pdftex]{graphicx}
\else
\fi

\usepackage{amsmath}
\usepackage{xcolor}
\usepackage{algorithm}
\usepackage{algorithmicx}
\usepackage{amsmath,algorithm,tabularx}
\usepackage{algpseudocode}
\usepackage{amssymb}
\usepackage{multirow}
\usepackage{makecell}
\usepackage{caption}
\usepackage{array}
\newcolumntype{P}[1]{>{\centering\arraybackslash}p{#1}}
\newcolumntype{C}{>{\centering\arraybackslash} m{6cm} }
\usepackage{url}
\usepackage[utf8]{inputenc}
\usepackage[english]{babel}

\usepackage[absolute]{textpos}
\makeatletter
\newcommand{\multiline}[1]{%
	\begin{tabularx}{\dimexpr\linewidth-\ALG@thistlm}[t]{@{}X@{}}
		#1
	\end{tabularx}
}
\makeatother

\begin{document}
\begin{textblock}{13}(1.5,0.35)
	\noindent S. Sun and H. Yan, ``Small-scale spatial-temporal correlation and degrees of freedom for reconfigurable intelligent surfaces," accepted to \textit{IEEE Wireless Communications Letters}, DOI:10.1109/LWC.2021.3112781.
\end{textblock}

\setlength{\abovedisplayskip}{2pt}
\setlength{\belowdisplayskip}{5pt}
	
\title{Small-Scale Spatial-Temporal Correlation and Degrees of Freedom for Reconfigurable Intelligent Surfaces}
\author{Shu Sun, \textit{member, IEEE}, and Hangsong Yan, \textit{member, IEEE}
\thanks{Shu Sun is with the Next Generation and Standards Group, Intel Corporation, Santa Clara, CA 95054, USA (e-mail: ss7152@nyu.edu).

Hangsong Yan is with the NYU WIRELESS research center and Tandon School of Engineering, New York University, Brooklyn, NY 11201, USA (e-mail: hy942@nyu.edu).}
}

\maketitle

\begin{abstract}
The reconfigurable intelligent surface (RIS) is an emerging promising candidate technology for \textcolor{black}{future} wireless networks, where the element spacing is usually of sub-wavelength. Only limited knowledge, however, has been gained about the spatial-temporal correlation behavior among the elements in an RIS. In this paper, we investigate the spatial-temporal correlation for \textcolor{black}{an RIS-enabled} wireless communication system. \textcolor{black}{Specifically, a} joint small-scale spatial-temporal correlation model is \textcolor{black}{derived} under isotropic scattering, which can be represented by a four-dimensional sinc function. Furthermore, \textcolor{black}{based upon} the spatial-only correlation at a certain time instant, \textcolor{black}{an essential RIS property -- the spatial degrees of freedom (DoF) -- is revisited, and} an analytical expression is \textcolor{black}{propounded} to characterize \textcolor{black}{the spatial DoF} for RISs with realistic hence \textcolor{black}{non-infinitesimal} element spacing and finite aperture sizes. \textcolor{black}{The results are vital to the accurate evaluation of various system performance metrics.}
\end{abstract}


\IEEEpeerreviewmaketitle

\section{Introduction}
\IEEEPARstart{M}{ASSIVE}  MIMO (multiple-input multiple-output)\cite{Marzetta10TWC} is one of the key enabling technologies for the fifth-generation (5G) wireless communications, which can bring tremendous advantages in spectral efficiency, energy efficiency, and power control\cite{Marzetta10TWC,Yan21IoTJ}. As a natural extension of Massive MIMO, more elements may be arranged in a small form factor if the element spacing \textcolor{black}{is further reduced} from the half-wavelength, so that the entire array can be regarded as a spatially-continuous electromagnetic aperture in its ultimate form\cite{Pizzo20JSAC}. This type of extended Massive MIMO is named Holographic MIMO\cite{Pizzo20JSAC,Huang20WC}. Meanwhile, analogous to the metasurface concept in the optical domain\cite{Holloway12APM,Sun_JOSA}, the sub-wavelength architecture has the potential to manipulate impinging electromagnetic waves through anomalous reflection, refraction, polarization transformation, among other functionalities, for wireless communication purposes. Therefore, such sort of structure is also referred to as reconfigurable intelligent surface (RIS), large intelligent surface, etc. In this paper, we use RIS as the blanket term for all the aforementioned two-dimensional (2D) sub-wavelength architectures. In order to unleash the full potentials of the RIS technology \textcolor{black}{which can find wide applications in both terrestrial and aerial communications \cite{Cao21JSAC}}, it is necessary to understand RIS's fundamental properties, among which the associated channel model is of paramount importance since it is the foundation of a multiplicity of aspects in wireless systems including deployment decision, algorithm selection, and performance evaluation\cite{Sun18TVT}.

Despite the upsurge in research interests of RIS, only limited work is available in the open literature on characterizing its channel model, especially the small-scale fading model. A parametric channel model for RIS-empowered systems has been presented in\cite{Basar20LATINCOM}, but without explicit and tractable expressions for the small-scale fading. The authors of \cite{Pizzo20JSAC} have studied the small-scale fading of RIS via wave propagation theories and established a Fourier plane-wave spectral representation of the three-dimensional (3D) stationary small-scale fading. In\cite{Bjornson20WCL}, a spatially-correlated Rayleigh fading model has been derived under isotropic scattering. Nevertheless, the investigation in\cite{Pizzo20JSAC} and \cite{Bjornson20WCL} did not consider the temporal correlation among RIS elements. Moreover, although the spatial degrees of freedom (DoF) for sufficiently dense and large RISs have been well studied\cite{Pizzo20JSAC}, the achievable DoF for more common cases with finite element spacing and aperture areas has received less attention. To fill in these gaps, this paper explores the joint spatial-temporal correlation models and DoF for realistic RISs under isotropic scattering. \textcolor{black}{The main contributions of this paper are two-fold: First, a closed-form four-dimensional (4D) sinc function is derived to describe the joint \textit{spatial-temporal} correlation, rather than the spatial-only correlation in the existing literature. Second, a simple but accurate analytical expression is proposed to quantify the spatial DoF for RISs with \textit{non-infinitesimal element spacing and limited apertures}, which has not been investigated in previous work to our best knowledge but is valuable to channel estimation and beamforming design for RISs\cite{Cao21JSAC,Bjornson20WCL,Sun21RIS}. Mathematical symbols and definitions are given in Table~\ref{tbl:symbol}.}
\begin{table}
	\caption{\textcolor{black}{Mathematical symbols and their definitions}}
	\label{tbl:symbol}
	\textcolor{black}{
	\begin{center}
		\begin{tabular}{|p{0.177\textwidth}||p{0.246\textwidth}|}
			\hline
			Symbol & Definition  \\ 
			\hline\hline
			Italicized letter, e.g. $a$ and $A$ & Scalar \\  \hline
			Bold lowercase letter, e.g., $\textbf{a}$ & Column vector \\ \hline
			Bold capital letter, e.g., $\textbf{A}$ & Matrix \\ \hline
			Superscript $T$ & Transpose of matrix or vector\\  \hline
			Superscript $H$ & Conjugate transpose of matrix or vector\\  \hline
			mod($a,b$) & Remainder after dividing $a$ by $b$\\  \hline
			$\lfloor a\rfloor$& Nearest integer no larger than $a$ \\  \hline
		\end{tabular}
	\vspace{-2.9em}
	\end{center}}
\end{table}

\section{System Model}
\begin{figure}
	\centering
	\includegraphics[width=0.5\columnwidth]{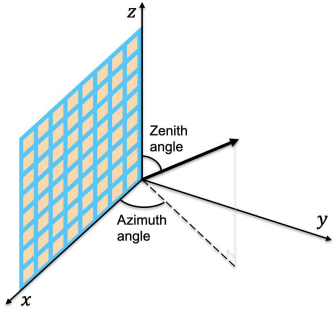}
	\caption{The orientation of an RIS and definitions of azimuth and zenith angles with respect to the associated coordinate system.}
	\label{fig:fig1}	
	\vspace{-1em}
\end{figure}
We consider an RIS equipped with $N$ elements in a wireless communication system, where the RIS can act as a transmitter, receiver, or interacting object in the propagation environment and can translate in the 3D space. For the purpose of exposition, we focus on the reception mode of the RIS in this paper, but the resultant spatial-temporal correlation models and DoF have more general applicability. The coordinate system and definitions of azimuth and zenith angles are aligned with \cite{38901}, as illustrated in Fig.~\ref{fig:fig1}. Without loss of generality, the RIS is assumed to be located in the $xz$ plane, where its horizontal and vertical lengths are $L_x$ and $L_z$, respectively. The motion direction of the RIS is described jointly by the azimuth angle $\varphi$ and zenith angle $\vartheta$ with a moving speed of $||\textbf{v}||$, where 

\vspace{-5pt}
\small
\begin{equation}\label{eq:v}
\textbf{v}=v[\text{cos}\varphi\text{sin}\vartheta,\text{sin}\varphi\text{sin}\vartheta,\text{cos}\vartheta]^T.
\vspace{-4pt}
\end{equation}
\normalsize

\noindent When there are $P$ plane waves impinging on the RIS at time instant $t$, the channel at the RIS can be expressed as 

\vspace{-0.1in}
\small
\begin{equation}\label{eq:hUR1}
\textbf{h}(t)=\sqrt{N/P}\sum_{p=1}^{P}\alpha_p\textbf{a}(\phi_p,\theta_p)e^{j\frac{2\pi}{\lambda}\textbf{v}^T\textbf{e}_pt}
\vspace{-0.1in}
\end{equation}
\normalsize

\noindent where $\alpha_p$ and $\phi_p (\theta_p)$ denote the complex gain and azimuth (zenith) angle of arrival of the $p$-th plane wave, respectively, $\lambda$ the wavelength, and $\textbf{e}_p=[\text{cos}\phi_p\text{sin}\theta_p,\text{sin}\phi_p\text{sin}\theta_p,\text{cos}\theta_p]^T$. Additionally, $\textbf{a}$ denotes the RIS array response vector

\vspace{-0.1in}
\small
\begin{equation}\label{eq:aR}
\begin{split}
\textbf{a}(\phi,\theta)=&1/\sqrt{N}\Big[1,...,e^{j\frac{2\pi}{\lambda}(x(n)d_x\text{cos}\phi\text{sin}\theta+z(n)d_z\text{cos}\theta)},...,\\
&e^{j\frac{2\pi}{\lambda}(x(N)d_x\text{cos}\phi\text{sin}\theta+z(N)d_z\text{cos}\theta)}\Big]^T
\end{split}
\end{equation}
\normalsize
\noindent where $d_x$ and $d_z$ are the adjacent element spacing of the RIS in the $x$ and $z$ directions, respectively. $x(n)$ and $z(n)$ are respectively the indices in the $x$ and $z$ directions for the $n$-th element, which are calculated as

\vspace{-10pt}
\small
\begin{equation}\label{eq:xz}
\begin{split}
x(n)=\text{mod}(n-1,N_x), z(n)=\lfloor(n-1)/N_x\rfloor
\end{split}
\end{equation}
\vspace{-15pt}
\normalsize

\noindent with $N_x$ standing for the number of elements per row. Denote the average attenuation of the $P$ plane waves as $\mu$, then as $P\to\infty$\footnote{\textcolor{black}{Note that although the coordinate system and definitions of azimuth and zenith angles in Fig.~\ref{fig:fig1} comply with the 3GPP TR 38.901\cite{38901}, the channel model in this paper is for isotropic scattering where the number of plane waves approaches infinity, which is also employed in the literature such as \cite{Pizzo20JSAC,Bjornson20WCL}. The DoF derived herein can be regarded as an upper bound among a variety of channel models. Non-isotropic-scattering channel models, such as those defined in \cite{38901}, are left for future work.}}, based on the central limit theorem, the normalized spatial-temporal correlation matrix \small$\textbf{R}(\tau)$ \normalsize is given by

\vspace{-6pt}
\small
\begin{equation}
\textbf{R}(\tau)=\frac{1}{\mu}\mathbb{E}\Big\{\textbf{h}(t)\textbf{h}^H(t+\tau)\Big\}.
\end{equation}
\vspace{-14pt}
\normalsize

\noindent Since $P\to\infty$, the discrete random variables $\phi_p$ and $\theta_p$ become continuous random variables $\phi$ and $\theta$, which are characterized by a certain angular distribution $f(\phi,\theta)$. Consequently, the ($m,n$)-th element of $\textbf{R}(\tau)$ can be expressed as in (\ref{eq:R1}). In the next section, we will conduct further explorations of $\textbf{R}(\tau)$ under isotropic scattering.
\begin{figure*}
	\small
	\begin{equation}\label{eq:R1}
	\begin{split}
	[\textbf{R}(\tau)]_{m,n}=&~N\mathbb{E}\Big\{[\textbf{a}(\phi,\theta)]_m[\textbf{a}^H(\phi,\theta)]_ne^{-j\frac{2\pi}{\lambda}\tau\textbf{v}^T\textbf{e}_p}\Big\}\\
	=&~\mathbb{E}\Big\{e^{j\frac{2\pi}{\lambda}[(x(m)-x(n))d_x\text{cos}\phi\text{sin}\theta+(z(m)-z(n))d_z\text{cos}\theta]}e^{-j\frac{2\pi}{\lambda}v\tau(\text{cos}\phi\text{sin}\theta\text{cos}\varphi\text{sin}\vartheta+\text{sin}\phi\text{sin}\theta\text{sin}\varphi\text{sin}\vartheta+\text{cos}\theta\text{cos}\vartheta)}\Big\}\\
	=&~\int_{0}^{\pi}\int_{0}^{\pi}\Big\{e^{j\frac{2\pi}{\lambda}[((x(m)-x(n))d_x-v\tau\text{cos}\varphi\text{sin}\vartheta)\text{cos}\phi\text{sin}\theta-v\tau\text{sin}\varphi\text{sin}\vartheta\text{sin}\phi\text{sin}\theta+((z(m)-z(n))d_z-v\tau\text{cos}\vartheta)\text{cos}\theta]}f(\phi,\theta)\Big\}d\phi d\theta,\\
	&~m,n=1,...,N
	\end{split}
	\end{equation}
	\rule{\textwidth}{0.4pt}
\end{figure*}
\vspace{-9pt}
\normalsize

\section{Joint Spatial-Temporal Correlation}
For the isotropic scattering environment, the angular distribution function has the following form

\vspace{-7pt}
\small
\begin{equation}
f(\phi,\theta)=\text{sin}(\theta)/(2\pi), ~\phi\in\big[0,\pi\big],\theta\in\big[0,\pi\big]
\end{equation}
\normalsize

\noindent thus (\ref{eq:R1}) can be recast as
\small
\begin{equation}\label{eq:R2}
\begin{split}
[\textbf{R}(\tau)]_{m,n}=&\frac{1}{2\pi}\int_{0}^{\pi}\int_{0}^{\pi}\Big\{\text{exp}\bigg(j\frac{2\pi}{\lambda}[((x(m)-x(n))d_x-\\
&v\tau\text{cos}\varphi\text{sin}\vartheta)\text{cos}\phi\text{sin}\theta-v\tau\text{sin}\varphi\text{sin}\vartheta\text{sin}\phi\text{sin}\theta+\\
&((z(m)-z(n))d_z-v\tau\text{cos}\vartheta)\text{cos}\theta]\bigg)\text{sin}\theta\Big\}d\phi d\theta,\\
&m,n=1,...,N
\end{split}
\end{equation}
\normalsize

\noindent \textbf{Proposition 1.} \textit{With isotropic scattering in the half-space in front of the RIS, the joint spatial-temporal correlation matrix $\textbf{R}(\tau)$ is given by} \textcolor{black}{(please see the Appendix for its proof)}

\vspace{-5pt}
\small
	\begin{equation}\label{eq:R3}
	\begin{split}
	[\textbf{R}(\tau)]_{m,n}=&\text{sinc}\bigg(\frac{2||\textbf{d}_m-\textbf{d}_n-\tau\textbf{v}||}{\lambda}\bigg),~m,n=1,...,N
	\end{split}
	\end{equation}
\normalsize

\noindent \textit{where $\text{sinc}(x)=\frac{\text{sin}(\pi x)}{\pi x}$ is the sinc function, $\textbf{d}_m$ and $\textbf{d}_n$ denote the coordinates of the $m$-th and $n$-th RIS element, respectively, and $\textbf{v}$ is provided in (\ref{eq:v}).}

As unveiled by Proposition 1, the joint spatial-temporal correlation among RIS elements is characterized by a 4D (the 3D space plus the time dimension) sinc function, from which the following observations can be made. First, the correlation is minimal only for some element spacing, instead of between any two elements, thus the i.i.d. Rayleigh fading model is not applicable in such a system, which is consistent with the conclusions made in \cite{Pizzo20JSAC,Bjornson20WCL}. Second, distinct from the situation in \cite{Bjornson20WCL} without temporal correlation statistics, the correlation in (\ref{eq:R3}) depends upon the spatial and temporal domains jointly. More specifically, the correlation is low when $||\textbf{d}_m-\textbf{d}_n-\tau\textbf{v}||$ is equal or close to integer multiples of half-wavelength, and reaches the maximum when $\textbf{d}_m-\textbf{d}_n=\tau\textbf{v}$, i.e., it is possible for two relatively distant elements to have high spatial-temporal correlation if the difference of their coordinate vectors aligns with the speed vector multiplied by the time interval. Physically, the $m$-th element arrives at the current location of the $n$-th element after traveling time interval $\tau$, hence it \grqq{sees}" the same channel as the $n$-th element at the current time instant. This observation is quite different from the spatial-only case where high correlation is usually yielded merely by closely-spaced elements. Additionally, the rank of $\textbf{R}(\tau=0)$ can be characterized by its non-trivial eigenvalues. The i.i.d. Rayleigh fading channel has non-trivial eigenvalues whose amount equals the number of antenna elements deployed, while the correlated channel has fewer dominant eigenvalues and smaller rank. Besides, if focusing on the correlation in the time domain only by setting $m=n$, then (\ref{eq:R3}) reduces to $\text{sinc}\Big(\frac{2||\tau\textbf{v}||}{\lambda}\Big)=\text{sinc}\big(\frac{2\tau v}{\lambda}\big)$, implying that the temporal correlation for a given RIS element is characterized by a one-dimensional sinc function which is independent of the motion direction. If defining the decorrelation time $\tau_\text{decor}$ as the time when the absolute correlation value drops to $1/e$ and remains within $1/e$ afterwards\cite{Sun17SmallScale}, then $\tau_\text{decor}\approx0.35\lambda/v$. Assuming $\lambda=0.1$ m (corresponding to 3 GHz carrier frequency) and $v=1$ m/s, then $\tau_\text{decor}\approx35$ ms, which is around 70 slots with a typical slot duration of 0.5 ms\cite{38211}, implying the need for infrequent channel information updates. 

\section{Spatial Degrees of Freedom}
\begin{figure}
	\centering
	\includegraphics[width=1.09\columnwidth]{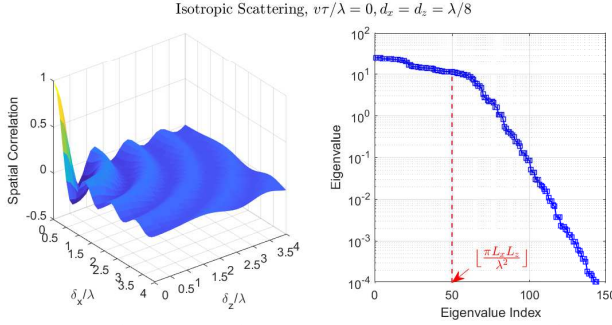}
	\caption{Spatial correlation among RIS elements (left) and eigenvalues of $\textbf{R}(\tau=0)$ in non-increasing order (right) under isotropic scattering, with $L_x=L_z=4\lambda$ and $d_x=d_z=\lambda/8$.}
	\label{fig:fig2}	
\end{figure}

As a special case of (\ref{eq:R3}), the spatial-only correlation can be attained by setting $v\tau/\lambda=0$. Fig.~\ref{fig:fig2} illustrates the small-scale spatial correlation among the RIS elements and the eigenvalues of $\textbf{R}(\tau=0)$ at a certain time instant with $L_x=L_z=4\lambda$ and $d_x=d_z=\lambda/8$, in which the left plot shows the pattern of the spatial-only sinc function, with $\delta_x$ and $\delta_z$ denoting respectively the horizontal and vertical distances in meters between RIS elements. It is evident from Fig.~\ref{fig:fig2} that the small-scale spatial correlation is high when the element spacing is noticeably smaller than the half-wavelength. If defining the decorrelation distance $d_\text{decor}$ in a similar way to the decorrelation time\cite{Sun17SmallScale}, then $d_\text{decor}\approx0.35\lambda$ in this case. Furthermore, as shown by the right plot of Fig.~\ref{fig:fig2}, the eigenvalues are highly unevenly distributed and the number of dominant eigenvalues is small compared to the total number of elements, unlike the i.i.d. Rayleigh fading channel. These observations corroborate the fact that the i.i.d. Rayleigh fading model should not be employed for the RIS even in an isotropic scattering environment\cite{Pizzo20JSAC,Bjornson20WCL}. In fact, the rank can be estimated as $\big\lfloor\frac{\pi L_xL_z}{\lambda^2}\big\rfloor$ for a sufficiently dense and large RIS\cite{Pizzo20JSAC}, which equals 50 in this case. The dominant $\big\lfloor\frac{\pi L_xL_z}{\lambda^2}\big\rfloor$ eigenvalues contain about $82\%$ of the total channel power herein.
\begin{figure}
	\centering
	\includegraphics[width=1.07\columnwidth]{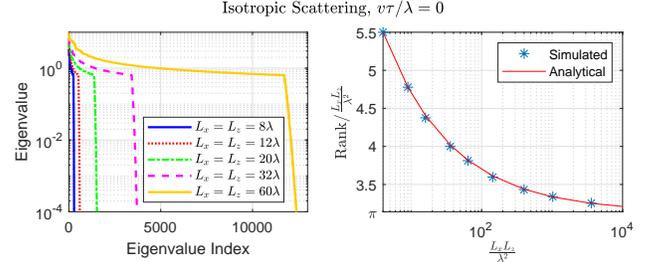}
	\caption{Eigenvalues of $\textbf{R}(\tau=0)$ in non-increasing order for various $L_x$ and $L_z$ (left), and ratios of rank of $\textbf{R}(\tau=0)$ to $\frac{L_xL_z}{\lambda^2}$ versus $\frac{L_xL_z}{\lambda^2}$ (right), with $L_x=L_z$ and $d_x=d_z=\lambda/2$.}
	\label{fig:rankVsArea}	
\end{figure}
\begin{figure}
	\centering
	\includegraphics[width=1.04\columnwidth]{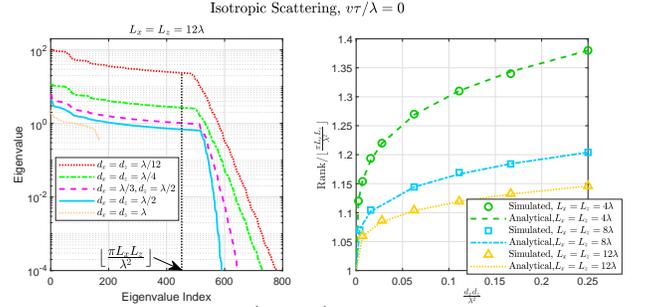}
	\caption{Eigenvalues of $\textbf{R}(\tau=0)$ in non-increasing order for various $d_x$ and $d_z$ with $L_x=L_z=12\lambda$ (left), and ratios of rank of $\textbf{R}(\tau=0)$ to $\lfloor\frac{\pi L_xL_z}{\lambda^2}\rfloor$ versus $\frac{d_xd_z}{\lambda^2}$ for various $L_x$ and $L_z$ (right).}
	\label{fig:fig3}	
	\vspace{-1em}
\end{figure}

It is worth noting from the right plot of Fig.~\ref{fig:fig2} that the actual number of dominant eigenvalues, i.e., the spatial DoF or rank beyond which the eigenvalues decay rapidly,  is perceptibly larger than $\big\lfloor\frac{\pi L_xL_z}{\lambda^2}\big\rfloor$ due to the finite size and element spacing of the underlying RIS. Since in practice an RIS usually has non-zero element spacing and non-infinite aperture (even with respect to the wavelength instead of physically), it is meaningful to investigate the actual spatial DoF and its relation with the approximation $\big\lfloor\frac{\pi L_xL_z}{\lambda^2}\big\rfloor$. To this end, we study the eigenvalues and rank $r$ of $\textbf{R}(\tau=0)$ for various $d_x,d_z,L_x$ and $L_z$. First, we examine $r$ for different square RIS\footnote{A square RIS has the lowest spatial DoF among all the rectangular RISs with the same area, thus the result here can serve as a lower bound on the DoF for different shapes of rectangular RISs. The upper bound corresponds to a column or row array occupying the same area.} aperture areas with a half-wavelength spacing. The left plot of Fig.~\ref{fig:rankVsArea} illustrates the eigenvalues for a wide range of $L_x$ and $L_z$, while the asterisk symbols in the right plot of Fig.~\ref{fig:rankVsArea} represent the ratio of the rank to RIS aperture size as a function of  the aperture size for square RISs, which shows that the ratio decreases with the aperture size and approaches $\pi$ for an infinitely large RIS. \textcolor{black}{According to the Lebesgue measure\cite[Eq. (1.2)]{Nowak05CMJ}, $r=\frac{\pi L_xL_z}{\lambda^2}+o(\frac{L_xL_z}{\lambda^2})$.} Inspired by this fact and based on simulation observations, we propose the following heuristic analytical expression to quantify the relation between $r$ and aperture size $\frac{L_xL_z}{\lambda^2}$

\vspace{-2pt}
\small
\begin{equation}\label{eq:Ratio1}
\begin{split} 
r=&~\textcolor{black}{\frac{\pi L_xL_z}{\lambda^2}+4.4\Big(\frac{L_xL_z}{\lambda^2}\Big)^{0.55}=}\frac{L_xL_z}{\lambda^2}\bigg(\pi+4.4\Big(\frac{L_xL_z}{\lambda^2}\Big)^{-0.45}\bigg),\\
&~L_x=L_z,d_x=d_z=\frac{\lambda}{2} \raisetag{1.2\baselineskip}
\end{split}
\end{equation}
\normalsize

\noindent \textcolor{black}{which} is represented by the solid curve in the right plot of Fig.~\ref{fig:rankVsArea} and matches the simulated values very well. \textcolor{black}{The following insights can be drawn from (\ref{eq:Ratio1}): First, the term $o(\frac{L_xL_z}{\lambda^2})$ in the Lebesgue measure turns out to be $4.4\big(\frac{L_xL_z}{\lambda^2}\big)^{0.55}$ herein, where $4.4$ and $0.55$ are obtained via numerical optimization, and it makes sense for the exponent to be smaller than $1$ since $o(\frac{L_xL_z}{\lambda^2})/\frac{\pi L_xL_z}{\lambda^2}$ should approach $0$ as $\frac{L_xL_z}{\lambda^2}\to\infty$. Second, $r/\frac{L_xL_z}{\lambda^2}$ approaches $\pi$ only if $\frac{L_xL_z}{\lambda^2}\to\infty$, while it is obviously greater than $\pi$ for finite $\frac{L_xL_z}{\lambda^2}$, since the theoretical limit $\frac{\pi L_xL_z}{\lambda^2}$ is derived only for infinitely large RISs hence the error increases as the RIS aperture decreases. Moreover, $r/\frac{L_xL_z}{\lambda^2}$ is larger than $\pi$ by the term $4.4\big(\frac{L_xL_z}{\lambda^2}\big)^{-0.45}$, i.e., $r/\frac{L_xL_z}{\lambda^2}$ decreases with $\frac{L_xL_z}{\lambda^2}$ at a rate proportional to $\frac{L_xL_z}{\lambda^2}$ raised to the power of $0.45$.}

Next, we study the rank for a given RIS aperture size with varying element spacing. The left plot of Fig.~\ref{fig:fig3} depicts the eigenvalues for a plurality of $d_x$ and $d_z$ with a given aperture size of $L_x=L_z=12\lambda$ as an example, where decreasing $d_x$ and/or $d_z$ is equivalent to increasing the number of elements $N$ at the RIS. The following key remarks can be made:
\vspace{-2.5pt}
\begin{itemize}[leftmargin=*]
	\item The spatial DoF (i.e., the number of dominant eigenvalues) does not increase with $N$ indefinitely, and on the contrary, it declines with $N$ when $d_x$ and/or $d_z$ becomes smaller than the half-wavelength. This phenomenon is likely ascribed to the growing spatial correlation among the RIS elements as $N$ increases, and the effect of this spatial correlation surpasses the potential extra DoF brought by the additional elements. 
	\item The theoretical limit $\big\lfloor\frac{\pi L_xL_z}{\lambda^2}\big\rfloor$ \textcolor{black}{for sufficiently dense and large RISs} acts as a lower bound on the spatial DoF when $d_x$ and $d_z$ are no larger than the half-wavelength, and is tight when $d_x$ and $d_z$ approach zero. 
	\item The absolute values of the eigenvalues increase with $N$ as expected, since a larger $N$ gives rise to higher array gain. 
\end{itemize}

\noindent It holds practical relevance to quantify the spatial DoF for realistic, non-infinite values of $d_x,d_z,L_x$ and $L_z$. Therefore, we now investigate the relation between the spatial DoF (i.e., rank of $\textbf{R}(\tau=0)$) and the theoretical limit $\big\lfloor\frac{\pi L_xL_z}{\lambda^2}\big\rfloor$ for different aperture sizes and element spacing of the RIS. The simulated values of the ratio $\rho$ of the rank to $\lfloor\frac{\pi L_xL_z}{\lambda^2}\rfloor$ versus $\frac{d_xd_z}{\lambda^2}$ are represented by discrete symbols in the right plot of Fig.~\ref{fig:fig3} for aperture sizes of $16\lambda^2,64\lambda^2$, and $144\lambda^2$, respectively, which reveals that $\rho$ varies with $\frac{d_xd_z}{\lambda^2}$ in a 2D parabolic-like manner for a given aperture size, and approaches $1$ as $\frac{d_xd_z}{\lambda^2}$ approximates $0$. We put forward the following heuristic formula  to compute the rank (i.e., spatial DoF)

\vspace{-8pt}
\small
\begin{equation}\label{eq:Ratio2}
\begin{split}
\text{Rank}=&\bigg\lfloor\frac{\pi L_xL_z}{\lambda^2}\bigg\rfloor\Bigg(1+\bigg(\frac{bd_xd_z}{\lambda^2}\bigg)^{\frac{1}{4}}\Bigg),~d_x\leq\frac{\lambda}{2},d_z\leq\frac{\lambda}{2}
\end{split}
\end{equation}
\vspace{-7pt}
\normalsize

\noindent where \textcolor{black}{the exponent $\frac{1}{4}$ stems from the joint parabolic terms $(d_x/\lambda)^{\frac{1}{2}}$ and $(d_z/\lambda)^{\frac{1}{2}}$.} The coefficient $b=4\big(r\big/\big\lfloor\pi L_xL_z/\lambda^2\big\rfloor-1\big)^4$ \textcolor{black}{relying} only on the aperture size with $r$ given by (\ref{eq:Ratio1}), \textcolor{black}{and is obtained by setting $\frac{d_xd_z}{\lambda^2}=\frac{1}{4}$ and $\text{Rank}=r$ in (\ref{eq:Ratio2}).} The accuracy of (\ref{eq:Ratio2}) is evaluated through comparison with the simulated results, shown in the right plot of Fig.~\ref{fig:fig3}, where the analytical values are limned by the dashed and/or dotted curves. It is evident from the right plot of Fig.~\ref{fig:fig3} that the analytical expression can well characterize the actual rank.
\vspace{-10pt}
\section{Numerical Results \textcolor{black}{for Spatial-Temporal Correlation}}
\begin{figure}
	\centering
	\includegraphics[width=0.9\columnwidth]{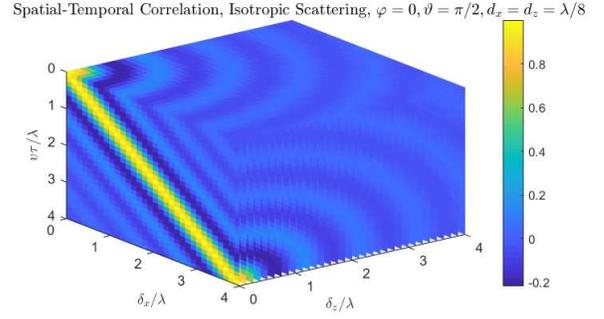}
	\caption{Joint spatial-temporal correlation among RIS elements under isotropic scattering. The RIS moving direction is along the $x$ axis.}
	\label{fig:fig4}	
\end{figure}

\begin{figure}
	\centering
	\includegraphics[width=0.9\columnwidth]{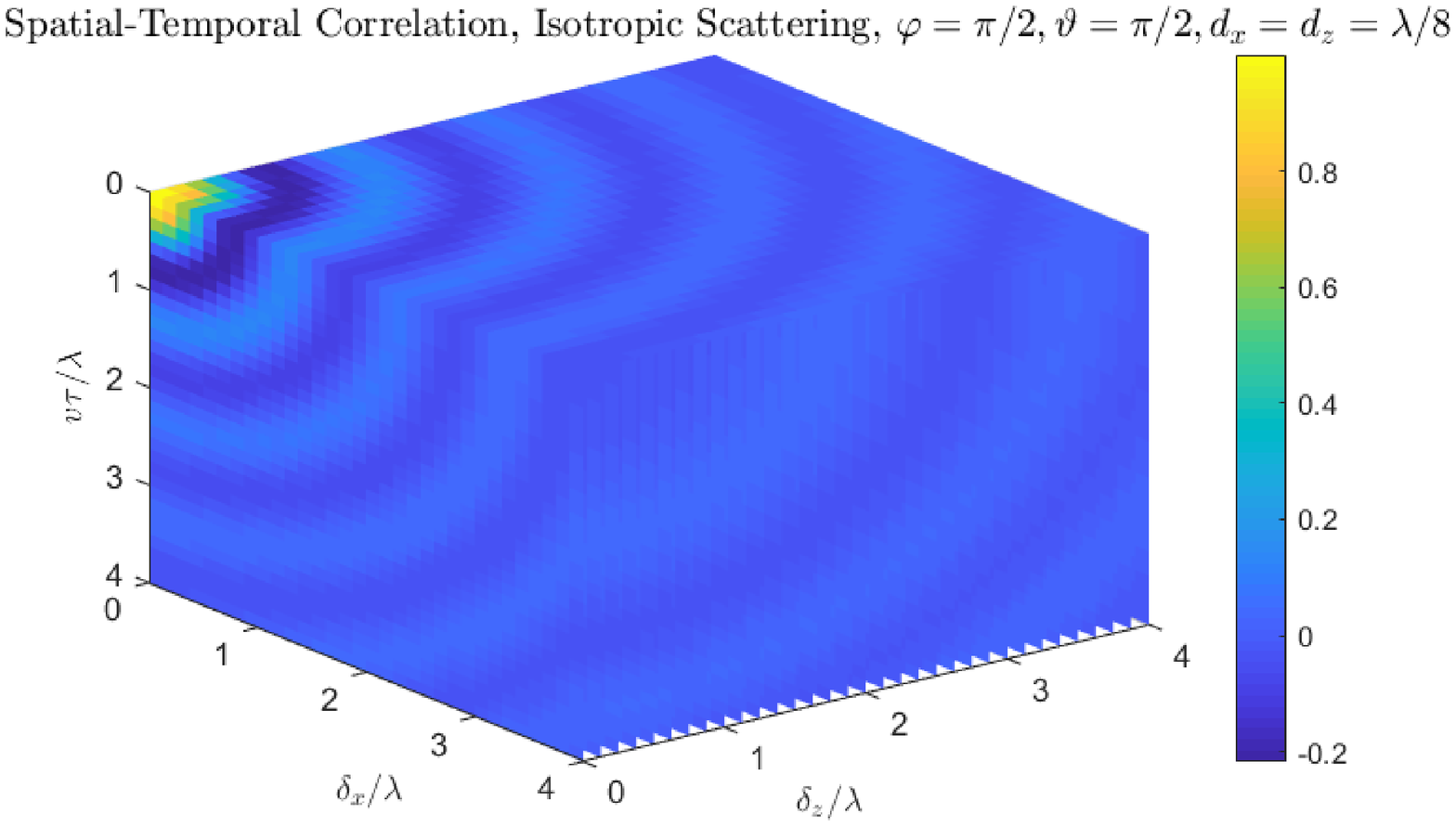}
	\caption{Joint spatial-temporal correlation among RIS elements under isotropic scattering. The RIS moving direction is along the $y$ axis.}
	\label{fig:fig5}	
\end{figure}

\begin{figure}
	\centering
	\includegraphics[width=0.9\columnwidth]{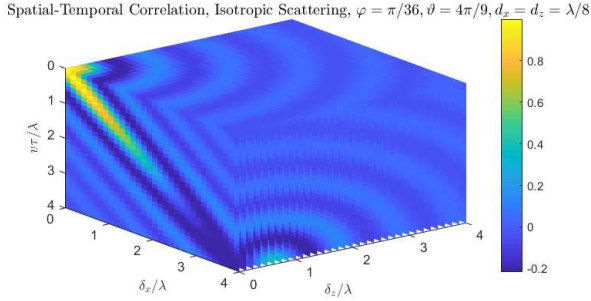}
	\caption{Joint spatial-temporal correlation among RIS elements under isotropic scattering. The RIS moving direction is $\varphi=\pi/36, \vartheta=4\pi/9$.}
	\label{fig:fig6}	
\end{figure}

Simulations are performed to more thoroughly inspect the small-scale spatial-temporal correlation behavior among the RIS elements. In the simulations, $L_x=L_z=4\lambda,d_x=d_z=\lambda/8$, so that $N_x=33,N=N_x^2$, unless otherwise specified. \textcolor{black}{The colorbars in Figs.~\ref{fig:fig4}-\ref{fig:fig6} represent the spatial-temporal correlation coefficient.}

The joint spatial-temporal correlation pattern when the RIS moves along the $x$ direction is depicted in Fig.~\ref{fig:fig4}, where the top horizontal slice manifests the spatial-only correlation equivalent to the left plot of Fig.~\ref{fig:fig2}, while the vertical slices delineate the joint spatial-temporal correlation along the $x$ direction for multiple samples at the $z$ direction. As shown by the vertical slice at $\delta_z/\lambda=0$, the strongest correlation occurs when $\delta_x/\lambda=v\tau/\lambda$, and generally decreases accompanied with oscillation, as precisely described by the sinc function in (\ref{eq:R3}). If collectively looking at the correlation pattern across multiple $z$ positions at $\delta_x/\lambda=4$, we can see that it is also sinc-like behavior which resembles that of the top horizontal slice, indicating the equivalence between spatial and temporal shifts, i.e., the correlation distribution at $\delta_x/\lambda=0$ and $v\tau/\lambda=0$ is equivalent to that at $\delta_x/\lambda=4$ and $v\tau/\lambda=4$ if the motion direction of the RIS aligns with the $x$ axis. Due to the rotational invariant resultant from the isotropy nature of the scattering, it is anticipated that the correlation pattern remains the same if the RIS moves along the $z$ direction with an exchange of $\delta_x/\lambda$ and $\delta_z/\lambda$ axes in Fig.~\ref{fig:fig4}.

Fig.~\ref{fig:fig5} displays the spatial-temporal correlation when the motion direction of the RIS is perpendicular to its surface. As evident from Fig.~\ref{fig:fig5}, the temporal correlation matches the spatial correlation if $v\tau/\lambda=\delta_x/\lambda$ or $v\tau/\lambda=\delta_z/\lambda$, in other words, there is no extra correlation induced by the movement of the RIS. This is because the moving direction is normal to the RIS hence incurring no correlation among the RIS elements. A more general case with the moving angles $\varphi=5^\circ,\vartheta=80^\circ$ is demonstrated in Fig.~\ref{fig:fig6}, which shows that the joint spatial-temporal correlation bears a resemblance to a weighted combination of the correlation patterns in Figs.~\ref{fig:fig4} and~\ref{fig:fig5}, since the motion direction can be regarded as a linear combination along the orthogonal directions in those two figures. \textcolor{black}{Particularly, the correlation in the $\delta_z/\lambda=0$ plane is weaker than that at the corresponding positions in Fig.~\ref{fig:fig4} but stronger than in Fig.~\ref{fig:fig5}, and gradually decreases along the $\delta_x/\lambda=v\tau/\lambda$ direction instead of remaining constant, attributed to the non-zero angle between the motion direction and the RIS facet.}

\vspace{-13.5pt}
\section{Conclusion}
\vspace{-3pt}
In this paper, we have derived a tractable analytical expression for the joint small-scale spatial-temporal correlation among the elements of a moving RIS under isotropic scattering, and a heuristic and accurate formula to characterize the achievable spatial DoF for square RISs with practical thus finite element spacing and aperture sizes. The joint spatial-temporal correlation can be modeled by a 4D sinc function. The contributions and observations in this paper can provide enlightenment on the proper exploitation of the RIS technology in the next-generation wireless communications. Future work will study the spatial-temporal correlation and other features of RISs under non-isotropic scattering.
\vspace{-6pt}
\section*{\textcolor{black}{Appendix}}
\vspace{-5pt}
\textcolor{black}{\textit{Proof for Proposition 1}: It is noteworthy that (\ref{eq:R2}) applies when the RIS is located in parallel with the $xz$ plane, so that the $y$ coordinate of any RIS element is the same hence canceled out when subtracting one from another. Nevertheless, in more general cases where the RIS is arbitrarily oriented, (\ref{eq:R2}) can be extended to}

\vspace{-10pt}
\small
\textcolor{black}{
\begin{equation}\label{eq:R4}
\begin{split}
[\textbf{R}(\tau)]_{m,n}=&1/(2\pi)\int_{0}^{\pi}\int_{0}^{\pi}\big\{\exp\big(j2\pi/\lambda[((x(m)-x(n))d_x-\\
&v\tau\text{cos}\varphi\text{sin}\vartheta)\text{cos}\phi\text{sin}\theta+((y(m)-y(n))d_y-\\
&v\tau\text{sin}\varphi\text{sin}\vartheta)\text{sin}\phi\text{sin}\theta+((z(m)-z(n))d_z-\\
&v\tau\text{cos}\vartheta)\text{cos}\theta]\big)\text{sin}\theta\big\}d\phi d\theta \raisetag{2.2\baselineskip}
\end{split}
\end{equation}}
\vspace{-9pt}
\normalsize

\noindent \textcolor{black}{where $y(n)$ is the index in the $y$ direction for the $n$-th element. Due to the isotropy of the scattering environment, the RIS can be rotated to result in a new set of coordinates $\{\tilde{x}(n),\tilde{y}(n),\tilde{z}(n),\tilde{d}_x,\tilde{d}_y,\tilde{d}_z\}$, based on which (\ref{eq:R4}) can be transformed to}

\vspace{-0.1in}
\small
\textcolor{black}{
\begin{equation}\label{eq:R5}
\begin{split}
[\textbf{R}(\tau)]_{m,n}=&1/(2\pi)\int_{0}^{\pi}\int_{0}^{\pi}\big\{\exp\big(j2\pi/\lambda[((\tilde{x}(m)-\tilde{x}(n))\tilde{d}_x-\\
&v\tau\text{cos}\varphi\text{sin}\vartheta)\text{cos}\phi\text{sin}\theta+((\tilde{y}(m)-\tilde{y}(n))\tilde{d}_y-\\
&v\tau\text{sin}\varphi\text{sin}\vartheta)\text{sin}\phi\text{sin}\theta+((\tilde{z}(m)-\tilde{z}(n))\tilde{d}_z-\\
&v\tau\text{cos}\vartheta)\text{cos}\theta]\big)\text{sin}\theta\big\}d\phi d\theta \raisetag{2.2\baselineskip}
\end{split}
\end{equation}}
\vspace{-7pt}
\normalsize

\noindent \textcolor{black}{The rotation angles can be selected such that $(\tilde{x}(m)-\tilde{x}(n))\tilde{d}_x-v\tau\text{cos}\varphi\text{sin}\vartheta=0$ and $(\tilde{y}(m)-\tilde{y}(n))\tilde{d}_y-v\tau\text{sin}\varphi\text{sin}\vartheta=0$, the expression in (\ref{eq:R5}) thus simplifies to}

\vspace{-0.1in}
\small
\textcolor{black}{
\begin{equation}\label{eq:R6}
\begin{split}
[\textbf{R}(\tau)]_{m,n}=&1/(2\pi)\int_{0}^{\pi}\int_{0}^{\pi}\big\{\exp\big(j2\pi/\lambda||\textbf{d}_m-\textbf{d}_n-\tau\textbf{v}||\text{cos}\theta]\big)\\
&\times\text{sin}\theta\big\}d\phi d\theta\\
\stackrel{(a)}{=}&\frac{\text{sin}\big(\frac{2\pi}{\lambda}||\textbf{d}_m-\textbf{d}_n-\tau\textbf{v}||\big)}{\frac{2\pi}{\lambda}||\textbf{d}_m-\textbf{d}_n-\tau\textbf{v}||} \raisetag{2.5\baselineskip}
\end{split}
\end{equation}}
\vspace{-5pt}
\normalsize

\noindent \textcolor{black}{where $(a)$ follows by employing Euler's formula.}

\ifCLASSOPTIONcaptionsoff
  \newpage
\fi
\vspace{- 6pt}
\bibliographystyle{IEEEtran}
\bibliography{Sun_WCL2021-1297}

\end{document}